\documentclass[twocolumn,showpacs,showkeys,prd]{revtex4}
\usepackage{graphicx}

\begin{document}
%%%%%%%%%%%%%%%%%%%%%%%%%%%%%%%%%%%%%%%%%%%%%%%%%%%%%%%%%%%%%%%%%%%%%%

\title{$0\nu2\beta$ decay and neutrino magnetic moment}

\author{Marek G\'o\'zd\'z} \email{mgozdz@kft.umcs.lublin.pl}
\author{Wies{\l}aw A. Kami\'nski}

\affiliation{
Department of Informatics, Maria Curie-Sk{\l}odowska University,
ul. Akademicka 9, 20-033 Lublin, Poland}

\begin{abstract}
  We propose a new channel of the neutrinoless double beta decay, based
  on the fact that in the presence of an external nonuniform magnetic
  field, the transition between neutrino and antineutrino can take place
  through the induced neutrino magnetic moment. We calculate the analog
  of the effective neutrino mass for this channel and show that, for
  certain values of the external magnetic field, a resonant enhancement
  can be obtained.
\end{abstract}

% 13.40.Em 	Electric and magnetic moments
% 11.30.Pb      Supersymmetry
% 14.60.Pq 	Neutrino mass and mixing
% 12.60.-i 	Models beyond the standard model
% 12.90.+b 	Miscellaneous theoretical ideas and models

\maketitle

%%%%%%%%%%%%%%%%%%%%%%%%%%%%%%%%%%%%%%%%%%%%%%%%%%%%%%%%%%%%%%%%%%%%%
\section{Introduction}
The neutrinoless double beta decay ($0\nu2\beta$) is a hypothetical
process in which a nucleus undergoes two beta transitions, while
neutrinos are not emitted, $(A,Z)\to (A,Z+2) + 2e$, $(A,Z)\to (A,Z-2) +
2\bar e$. This decay violates the electron lepton number by two units
and we know from neutrino oscillations that the flavor lepton numbers
are not conserved, in general. The $0\nu2\beta$ decay also violates the
total lepton number $\Delta L=2$; therefore, it requires some mechanism
which allows for the total lepton number to be broken.

The observation of the $0\nu2\beta$ decay will also prove that neutrinos
are Majorana particles \cite{schechter}, regardless of the underlying
nonstandard mechanism driving this process. It is therefore of great
importance to study this problem.

It is well known that neutrinos may undergo resonant flavor transitions
when propagating through matter \cite{MSW}. It has also been shown
\cite{B,aps93-plb,aps93-prd} that the additional presence of a
nonuniform magnetic field may induce transitions between neutrinos of
different helicities. Thus, to obtain the Pontecorvo oscillations
$\nu_{\alpha} \to \bar \nu_{\alpha}$, where $\alpha=e,\mu,\tau$, a
two-step process is necessary. This process will be driven by the
interplay between the neutrino mixing matrix $U$ and the neutrino
magnetic moment $\mu_{\alpha\beta}$.

The only magnetic moment for neutral particles can be induced in the
second-order processes, and for neutrinos in the standard model (SM)
this means a one-loop process defined by the interaction vertices of the
form $\nu\to W+e$, which conserve the electrical charge and lepton
number \cite{Kayser}. The induced neutrino magnetic moment in the SM has
been estimated to be $3.2 \times 10^{-19} (m_\nu)\mu_B$
\cite{FujikawaShrock}, where $m_\nu$ is the neutrino mass and $\mu_B =
e/(2m_e)$ is the Bohr magneton. For the neutrino mass $m_\nu=0.05$~eV
this yields $1.6 \times 10^{-20} \mu_B$ \cite{Shrock}. Going beyond the
standard model, e.g., introducing supersymmetry, it is possible to
enhance its value \cite{th-magmom1,th-magmom2}; however, experimental
bounds point towards values not greater than $\approx 10^{-11}\mu_B$
\cite{ex-magmom}.

In the following we discuss the process of neutrino-antineutrino
oscillations in nuclear matter with a nonuniform external magnetic
field. We show that this process may be resonantly enhanced, boosting
the half-life of the neutrinoless double beta decay.

\section{Corrections to the neutrino mass}

The presence of matter changes the energy levels of neutrinos through
neutral and charged current reactions. The neutral current potential is
given, in general, by (see, e.g., Ref.~\cite{mpal})
\begin{equation}
  V_{nc} = \sqrt{2} G_F \sum_{f=e,n,p} n_f \left(
    I_3^{(f)} -2 q^{(f)} \sin^2\theta_W \right),
\label{eq:Vnc}
\end{equation}
where $G_F$ is the Fermi constant, $\theta_W$ the Weinberg angle, $I_3$
the third component of the weak isospin, $q$ the electric charge, and
$n_f$ denotes the number densities of the fermions. In the
specific case of nuclear matter, the electrons are absent, $n_e=0$, and
the contributions come from the nucleons only. As for the charged
current interactions, they come from interactions between electron (mu,
tau) neutrinos and electrons (mu, tau leptons), so in our case they will
be absent as well. We conclude therefore that nuclear matter shifts the
overall scale of the neutrino energy levels but does not distinguish
between flavors or chiralities.

The presence of an external magnetic field, however, distinguishes
neutrinos from antineutrinos. Let us denote by $B=B(t)$ the component of
the magnetic field projected on the plane perpendicular to the neutrino
momentum. The direction of $B(t)$ is described by the angle $\phi(t)$,
and by $\dot\phi = d\phi(t)/dt$ we denote its time derivative. In what
follows we summarize the main points that lead to neutrino-antineutrino
transition, as they were discussed in
Refs.~\cite{aps93-plb,aps93-prd}. First, we switch to the coordinate
system which rotates together with $B$. In such a situation the energy
correction to the spin $-1/2$ neutrinos will be $+\dot\phi/2$, while for
the spin $+1/2$ antineutrinos, it will be $-\dot\phi/2$.

Assuming that we have only two flavors of neutrinos and working in
the basis
\begin{equation}
  (\nu_e,\nu_\mu,\bar\nu_e,\bar\nu_\mu)^T,
\end{equation}
the mixing matrix depends on one vacuum mixing angle only:
\begin{equation}
  U = \left(
  \begin{array}{cc} \
    \cos\theta & \sin\theta \\ -\sin\theta & \cos\theta
  \end{array} \right ).
\end{equation}
Furthermore one can write the Hamiltonian in the general form
\begin{equation}
  H = \left(
    \begin{array}{cc} \
      H_\nu & [B\mu] \\ -[B\mu] & H_{\bar\nu}
    \end{array} \right ),
\label{eq:H-2nu}
\end{equation}
where the neutrino and antineutrino blocks are given by
\begin{eqnarray}
  H_\nu &=& 
  E + V_{nc} + \frac{1}{2} \dot\phi + \frac{1}{2E} {\cal M}, \\
  H_{\bar\nu} &=& 
  E + V_{nc} - \frac{1}{2} \dot\phi + \frac{1}{2E} {\cal M}, 
\end{eqnarray}
with
\begin{eqnarray}
  {\cal M} = \left(
    \begin{array}{cc}
      m_1^2\cos^2\theta + m_2^2\sin^2\theta & 
      \frac{\Delta m^2}{2}\sin2\theta \\ 
      \frac{\Delta m^2}{2}\sin2\theta &
      m_1^2\sin^2\theta + m_2^2\cos^2\theta
    \end{array} \right ),
\label{eq:M}
\end{eqnarray}
and $\Delta m^2 = m_2^2-m_1^2$. To obtain
Eqs.~(\ref{eq:H-2nu})--(\ref{eq:M}) we have assumed that the energies of
the mass eigenstate neutrinos $E_i$ are $E_i = p + m_i^2/(2p)$ in the
relativistic case, for which the average neutrino energy $E$ is
approximately equal to its momentum $p$.

The off-diagonal blocks are
\begin{equation}
  [B\mu] = \left(\begin{array}{cc} 
      0 & B\mu_{e\mu} \\ -B\mu_{e\mu} & 0 
    \end{array} \right ),
\end{equation}
where $\mu_{e\mu}$ is the Majorana neutrino transition magnetic
moment. For Majorana neutrinos the magnetic moment is antisymmetric;
therefore, transitions between same flavors are forbidden.

After the diagonalization of the Hamiltonian (\ref{eq:H-2nu}) one finds
its eigenvalues in the form
\begin{equation}
  E + V_{nc} + \frac{m'^2}{2E},
\end{equation}
where the corrected neutrino masses are
\begin{eqnarray}
\label{eq:2n-mass}
  && m'^2 = \\
  && \frac{1}{2} \left(
    m_1^2 + m_2^2 \pm
    \sqrt{(4EB\mu_{e\mu})^2 + (E\dot\phi\pm\Delta m^2)^2}
  \right). \nonumber
\end{eqnarray}
In this process the neutrino is a virtual particle; as such it is not on
its mass shell, so it is not bound by the relation $E^2 = p^2 + m^2$. We
identify the small splitting induced by the $E\dot\phi$ term with the
neutrino-antineutrino masses, while the big splitting defines the first
and second mass eigenstates. One sees clearly that in the absence of the
$\dot\phi$ term, the neutrinos and antineutrinos will have degenerate
masses.

%%%%%%%%%%%%%%%%%%%%%%%%%%%%%%%%%%%%%%%%%%%%%%%%%%%%%%%%%%%%%%%%%%%%%%
\section{Neutrinoless double beta decay rate}

\begin{figure}
  \centering
  \includegraphics[width=0.9\columnwidth]{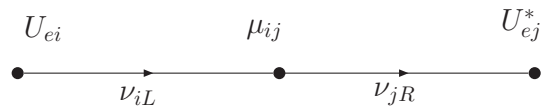}
  \caption{Internal line of the Feynman diagram describing the
    $0\nu2\beta$ decay driven by the magnetic moment of the neutrino.}
  \label{fig1}
\end{figure}

We are interested in the process in which the electron neutrino emitted
in one beta vertex is being absorbed as an electron antineutrino in the
second beta vertex. This will describe, apart from the nuclear matrix
element, the $0\nu2\beta$ process. The internal line of a Feynman
diagram in question is shown in Fig.~\ref{fig1}.

One has to notice first that neutrinos are produced and absorbed as
flavor eigenstates (electron neutrinos) in each beta vertex, but they
propagate through matter as mass eigenstates $\nu_i$. Therefore, the
magnetic moment $\mu_{ij}$ connects neutrino mass eigenstate $i$ with an
antineutrino mass eigenstate $j$ and is given by
\begin{equation}
  \mu_{ij} = \sum_{\alpha,\beta} U_{\alpha i}^* \mu_{\alpha\beta} U_{\beta j}.
\end{equation}
The propagation between two beta vertices, as depicted in
Fig.~\ref{fig1}, is then described by
\begin{eqnarray}
  \label{eq:chi1}
  && \chi = \sum_{\alpha,\beta} \mu_{\alpha\beta} \times \\
  &&
  \left( 
    \sum_{i=1,2} 
    U_{ei} \frac{\not p + \bar m'_i}{p^2-\bar m'^2_i} U_{\alpha i}^*
  \right ) \left (
    \sum_{j=1,2} 
    U_{\beta j}^* \frac{\not p + m'_j}{p^2-m'^2_j} U_{e j}^*
  \right ), \nonumber
\end{eqnarray}
where $p$ is the momentum exchanged between the beta vertices. In the
two-neutrino case we have explicitly
\begin{equation}
  \chi = \mu_{e\mu}
  \sum_{i,j} \left(
  \frac{U_{ei} U_{ei}^*}{\not p - \bar m'_i}
  \frac{U_{\mu j} U_{ej}^*}{\not p -  m'_j} -
  \frac{U_{ei} U_{\mu i}^*}{\not p - \bar m'_i}
  \frac{U_{e j} U_{ej}^*}{\not p -  m'_j}
  \right ),
\label{eq:chi2}
\end{equation}
where the antisymmetric property of the magnetic moment has been
used. We point out that a nonzero $\chi$ can have two sources. The
difference in effective masses of neutrinos and antineutrinos, $m'_i
\not = \bar m'_i$, is the first and most obvious one. However, even for
degenerate masses, a nonzero $\chi$ appears if the $U$ matrix is not
unitary. The general form of the neutrino mixing matrix contains complex
phases which describe possible $CP$ violation in neutrino
oscillations. These phases are unknown, although recent best-fit values
\cite{tortola} suggest a small $CP$ violation present in $U$.

The factor $\chi$ is related to the half-life of the $0\nu2\beta$ decay
through
\begin{equation}
  \left( T_{1/2}^{0\nu} \right)^{-1} = 
  G^{0\nu} |M^{0\nu}|^2 |B\chi|^2,
  \label{eq:T12}
\end{equation}
where $G^{0\nu}$ is the phase space factor and $M^{0\nu}$ is the nuclear
matrix element calculated within a certain (approximate) nuclear model.

It follows from Eq.~(\ref{eq:chi2}) that the factor $\chi$ is resonantly
enhanced around the poles in the denominators. The equality $m=p\approx
E$ yields, from Eq.~(\ref{eq:2n-mass}),
\begin{equation}
  \label{eq:phi}
  E\dot\phi = \pm \sqrt{(2E^2-(m_1^2+m_2^2))^2 -
    (4EB\mu_{e\mu})^2} \pm \Delta m^2.
\end{equation}
As the magnetic field corrections are energy dependent, and the typical
energy exceeds the mass eigenvalues by many orders of magnitude, the
neutrino masses in Eq.~(\ref{eq:2n-mass}) are dominated by the factors
proportional to $E$. The average momentum transfer, which for
relativistic neutrinos is comparable to the total energy $E$, in a
neutrinoless double beta decay is assessed from the nuclear radius and
usually taken to be $\sim 100$~MeV. The magnetic field strength $B$ can
be realistically estimated to be of the order of ${\cal O}(1)$~T. The
magnetic moment, on the other hand, does not exceed $10^{-11}\mu_B$
\cite{ex-magmom}. Neutrino masses squared $m_1^2$ and $m_2^2$ are of the
order of $10^{-5}$~eV \cite{tortola}. Taking all of this into account,
one finally arrives at the relation
\begin{equation}
  \dot\phi \approx 2E,
\end{equation}
which defines, to a good approximation, the resonance condition. In the
typical case of $E=100$~MeV, this yields $\dot\phi \sim 10^{23}$~Hz.

%%%%%%%%%%%%%%%%%%%%%%%%%%%%%%%%%%%%%%%%%%%%%%%%%%%%%%%%%%%%%%%%%%%%%%
\section{Summary}

Neutrino propagating in a nonuniform magnetic field may undergo
Pontecorvo transition to a $CP$-conjugated state of the same flavor, as
it was shown in \cite{aps93-plb,aps93-prd}. In this communication we
have pointed out that this process may mediate a neutrinoless double
beta decay. The difference between this and previously discussed
channels, like the mass mechanism, pion mechanism, sparticle mediation
and others, is that a resonancelike dependence on the frequency of the
external magnetic field exists. Thus, in principle, it may be possible
to enhance the rate of the $0\nu2\beta$ decay by controlling the field.

One should notice that when the resonance condition is met, the
propagator of a certain neutrino mass eigenstate is boosted. This means
that also the usual mass mechanism of the $0\nu2\beta$ decay should also
be enhanced. Being a higher order process, the magnetic moment channel
seems to be subdominant to the mass channel. However, one can also tune
the magnetic field to resonantly enhance the propagation of the
antineutrino mass eigenstate. In this case the magnetic moment channel
should dominate all others. A more detailed discussion of this problem
will appear in an upcoming paper. At present, we are not aware of any
experimental possibilities to discriminate between these two cases.

We have discussed a simplified two-neutrino case in which the third mass
eigenstate is decoupled from the electron neutrino. A full discussion of
the realistic three-neutrino case together with a more thorough
numerical analysis will be presented in a subsequent paper.

%%%%%%%%%%%%%%%%%%%%%%%%%%%%%%%%%%%%%%%%%%%%%%%%%%%%%%%%%%%%%%%%%%%%%%

\section*{Acknowledgments}

This work has been financed by the Polish National Science Centre under
the decision number DEC-2011/01/B/ST2/05932.

%%%%%%%%%%%%%%%%%%%%%%%%%%%%%%%%%%%%%%%%%%%%%%%%%%%%%%%%%%%%%%%%%%%%


\begin{thebibliography}{99}

\bibitem{schechter} J. Schechter, J. W. F. Valle, Phys. Rev. D {\bf 25},
  2951 (1982).

\bibitem{MSW} L. Wolfenstein, Phys. Rev. D {\bf 17}, 2369 (1978);
  S. P. Mikheyev, A. Yu. Smirnov, Il Nuovo Cimento C {\bf 9}, 17 (1986).

\bibitem{B} A. Yu. Smirnov, Phys. Lett. B {\bf 260}, 161 (1991).

\bibitem{aps93-plb} E. Kh. Akhmedov, S. T. Petcov, A. Yu. Smirnov,
  Phys. Lett. B {\bf 309}, 95 (1993).

\bibitem{aps93-prd} E. Kh. Akhmedov, S. T. Petcov, A. Yu. Smirnov,
  Phys. Rev. D {\bf 48}, 2167 (1993).

\bibitem{Kayser} B. Kayser, Phys. Rev. D {\bf 26}, 1662 (1982).

\bibitem{FujikawaShrock} K. Fujikawa, R. E. Shrock,
  Phys. Rev. Lett. {\bf 45}, 963 (1980).

\bibitem{Shrock} R. E. Shrock, private communication.

\bibitem{th-magmom1} 
  W. J. Marciano, A. I. Sanda, Phys. Lett. B {\bf 67} 303 (1977); 
  B. W. Lee, R. E. Shrock, Phys. Rev. D {\bf 16} 1444  (1977); 
  J. Schechter, J. W. F. Valle, Phys. Rev. D {\bf 24} 1883 (1981);
  J. Schechter, J. W. F. Valle, Phys. Rev. D {\bf 25} 283  (1982);
  J. F. Nieves, Phys. Rev. D {\bf 26} 3152 (1982);
  R. E. Schrock, Nucl. Phys. B {\bf 206} 359 (1982);
  L. F. Li, F. Wilczek, Phys. Rev. D {\bf 25} 143 (1982).

\bibitem{th-magmom2}
  M. G\'o\'zd\'z, W. A. Kami\'nski, F. \v Simkovic, A. Faessler,
  Phys. Rev. D {\bf 74} 055007  (2006);
  M. G\'o\'zd\'z, W. A. Kami\'nski, F. \v Simkovic, 
  Int. J. Mod. Phys. E {\bf 15} 441 (2006);
  M. G\'o\'zd\'z, W. A. Kami\'nski, Phys. Rev. D {\bf 78} 075021 (2008);
  M. G\'o\'zd\'z, W. A. Kami\'nski, Int. J. Mod. Phys. E {\bf 18} 109 (2009);
  M. G\'o\'zd\'z, W. A. Kami\'nski, Int. J. Mod. Phys. E {\bf 19} 692 (2010);
  M. G\'o\'zd\'z, Phys. Rev. D {\bf 85} 055016 (2012).

\bibitem{ex-magmom} 
  Z. Daraktchieva {\it et al.} (MUNU Collaboration), 
  Phys. Lett. B {\bf 564} 190 (2003);
  Z. Daraktchieva {\it et al.} (MUNU Collaboration),
  Phys. Lett. B {\bf 615} 153 (2005);
  H.B. Li {\it et al.} (TEXONO Collaboration), 
  Phys. Rev. Lett. {\bf 90} 131802 (2003);
  G. Bellini {\it et al.} (Borexino Collaboration), 
  Phys. Lett. B {\bf 696} 191  (2011).

\bibitem{mpal} R. N. Mohapatra, P. B. Pal, {\it Massive Neutrinos in
    Physics and Astrophysics} (World Scientific, Singapore, 2004).

\bibitem{tortola} M. T\'ortola, Fortschr. Phys. {\bf 61}, 427 (2013).

\end{thebibliography}
\end{document}